%% file: main.tex
\documentclass[sigconf,nonacm]{acmart}
\usepackage{enumitem}
\usepackage{cleveref}
\usepackage{csquotes}
\usepackage{csvsimple}

\setcopyright{rightsretained}
\copyrightyear{2021}
\acmYear{2021}
\acmDOI{}

\acmPrice{}
\acmISBN{}

\title{Ontology drift is a challenge for explainable data governance}

\author{Jiahao Chen}
\email{jiahao.chen@jpmorgan.com}
\orcid{0000-0002-4357-6574}
\affiliation{%
  \institution{J.\ P.\ Morgan AI Research}
  \streetaddress{383 Madison Avenue}
  \city{New York}
  \state{New York}
  \country{USA}
  \postcode{10179-0001}
}

\begin{document}

\begin{abstract}
We introduce the needs for explainable AI that arise from 
Standard No.\ 239 from the Basel Committee on Banking Standards (BCBS 239),
which outlines 11 principles for  
effective risk data aggregation and risk reporting for financial institutions.
Of these, explainable AI is necessary for compliance in two key aspects:
data quality, and appropriate reporting for multiple stakeholders.
We describe the implementation challenges
for one specific regulatory requirement:
that of having a complete data taxonomy that is appropriate for firmwide use.
The constantly evolving nature of financial ontologies necessitate a continuous
updating process to ensure ongoing compliance.
\end{abstract}

\maketitle

\section{Compliance with the BCBS 239 standard requires explainable AI/ML}

Financial services companies have unique, industry-specific needs for explainable AI/ML,
many of which are driven in part by compliance needs \cite[Ch.\ 7]{wef2020}.
Some previous work describes how explainability needs arise from fairness considerations in specific business decisions like credit decisions \cite{Chen2018},
and other needs relate to regulatory requirements such as model risk management \cite{Chen2020b}.
Other needs for explainability in financial services have been covered by multiple workshop papers in the NeurIPS Workshops on AI in Financial Services (2018 and 2019).

\begin{figure}
\centering
\includegraphics[width=\columnwidth,keepaspectratio]{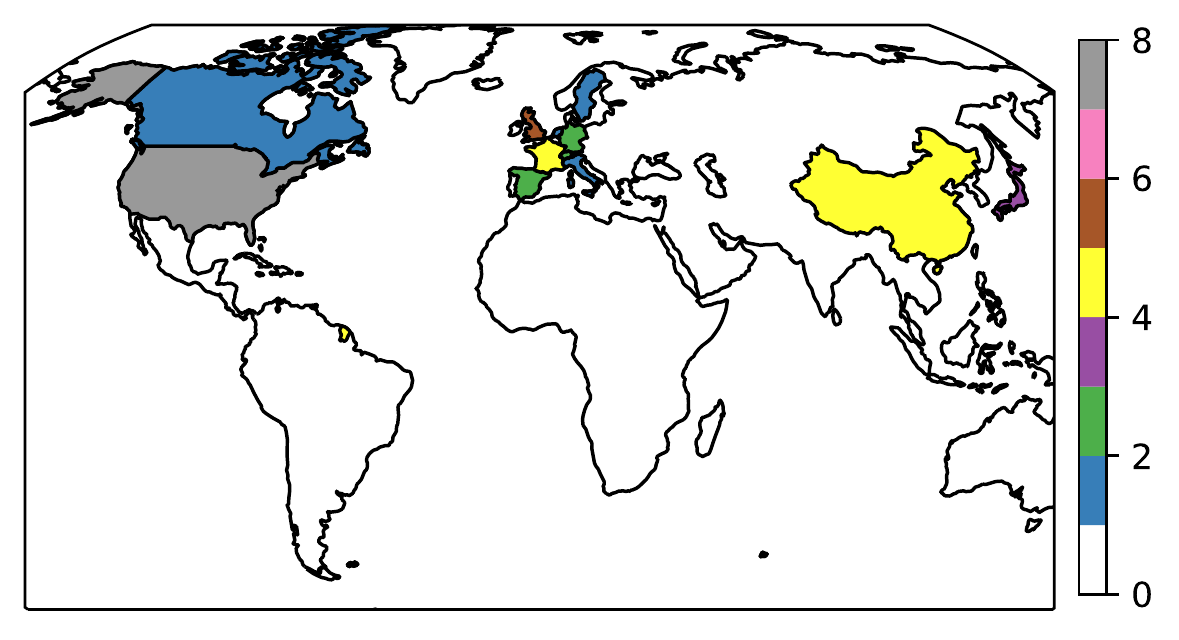}
\caption{Worldwide distributions of Global Systemically Important Banks (G-SIBs) in 2019 \cite{bcbs239pr}.}
\label{fig:gsibs-map}
\end{figure}

In this survey,
we describe how verifying compliance with a specific standard, No.\ 239 from the \cite{bcbs239}, drives specific explainability needs for Global Systemically Important Banks (G-SIBs, whose distribution worldwide is shown in \Cref{fig:gsibs-map}),
and presents opportunities for using explainable AI to service these compliance needs.
we do not confine ourselves to the narrow scope of variable level explanations that have been in vogue recently,
but rather take a more holistic view of what explanation means \cite{xaitutorial2020}.

\subsection{The 11 principles of BCBS 239}

The eleven principles for effective risk
data aggregation and risk
reporting for banks, as defined in BCBS 239,
are grouped into three topics as follows:

\renewcommand{\labelenumi}{\Roman{enumi}.}
\renewcommand{\labelenumii}{P\arabic{enumii}.}
\begin{enumerate}

\item Overarching governance and infrastructure

\begin{enumerate}[series=innerlist]
\item Governance [*]
\item Data architecture and IT infrastructure [*]
\end{enumerate}

\item Risk data aggregation capabilities

\begin{enumerate}[resume=innerlist]
\item Accuracy and integrity  [**]
\item Completeness [*]
\item Timeliness
\item Adaptability [*]
\end{enumerate}

\item Risk reporting practices

\begin{enumerate}[resume=innerlist]
\item Accuracy [**]
\item Comprehensiveness [*]
\item Clarity and usefulness [**]
\item Frequency [*]
\item Distribution
\end{enumerate}
\end{enumerate}

\noindent
A further three principles are specific to bank supervisors (regulators) only, and are not listed above.
The star ratings indicate our subjective evaluation as to whether a particular principle relates in a major [**] or minor [*] way to explainability requirements, which are described below.

\subsection{The explainability needs inherent in BCBS 239}%
\label{sec:explanation-needs}

We now describe how the BCBS standard can be interpreted (non-exhaustively)
as highlighting inherent needs for explainable AI.
Here, we ignore requirements for process documentation, automation, and infrastructure,
assuming that they are necessary prerequisites for explainable AI.
Relevant principles and paragraphs of the BCBS 239 standard are cross-referenced and paraphrased below.

We collect the explainability needs around input data and output reporting as follows:

\begin{enumerate}[label=(\Alph*)]
\item 
Input data should be of high quality.

\begin{enumerate}[label=(A\arabic*)]
\item
\textit{Data completeness.}
Verify that all relevant data is collected across the entire bank, across organizational boundaries, even as the organizational structure changes. (P1:29, P4:41--43)

\item
\textit{Data provenance.}
Trace the provenance of all data back to unique, authoritative sources. (P3:36(c,d))

\item
\textit{Quality control.}
Monitor data quality and have mitigation strategies for bad data. (P3:40, P7:53)

\item
\textit{Data description.}
Use a single, complete inventory of data, with a data dictionary and taxonomy describing data characteristics (metadata). (P2:33, P3:37, P9:67)
\end{enumerate}

\item 
Output reports should address the needs of multiple stakeholders.

\begin{enumerate}[label=(B\arabic*)]

\item
\textit{Flexible aggregation.}
Data aggregation should be customizable to user needs, and adaptable to both internal organization changes and extrinsic changes such as new regulations. (P6:49(b--d))

\item
\textit{Business needs.} 
Banks should justify their own reporting requirements based on their portfolios of business needs (P7:56, P8:59)

\item
\textit{Summarization level.}
Reports should be accurate, concise, comprehensive, and understandable to their intended recipients. (P7, P9)
Reports should be tailored to the needs of multiple stakeholders at different levels across the organization, such as the board, senior management, and risk committees, as well as external stakeholders like bank supervisors (regulators). (P9:61--66,68)
Reports to senior stakeholders should be more highly summarized. (P9:69)

\item
\textit{Reporting frequency,}
which should be appropriate for the stakeholder's needs. (P10:70)

\end{enumerate}

\end{enumerate}

At the high level, note that the data quality needs correspond roughly to P3,
whereas reporting needs correspond roughly to P7 and P9.
However, secondary requirements related to explainability are scattered throughout many of the other eleven principles codified in BCBS 239.
we therefore designate Principles P3, P7 and P9 as of major relevance to explainable AI,
and other principles mentioned above as of minor relevance.
In the next section, we will also show quantitative evidence that these principles are also those that demonstrate the least statistical evidence for compliance improvement.

\subsection{Compliance has been slower than expected}

\begin{figure*}
\centering
\includegraphics[width=0.33\textwidth,keepaspectratio]{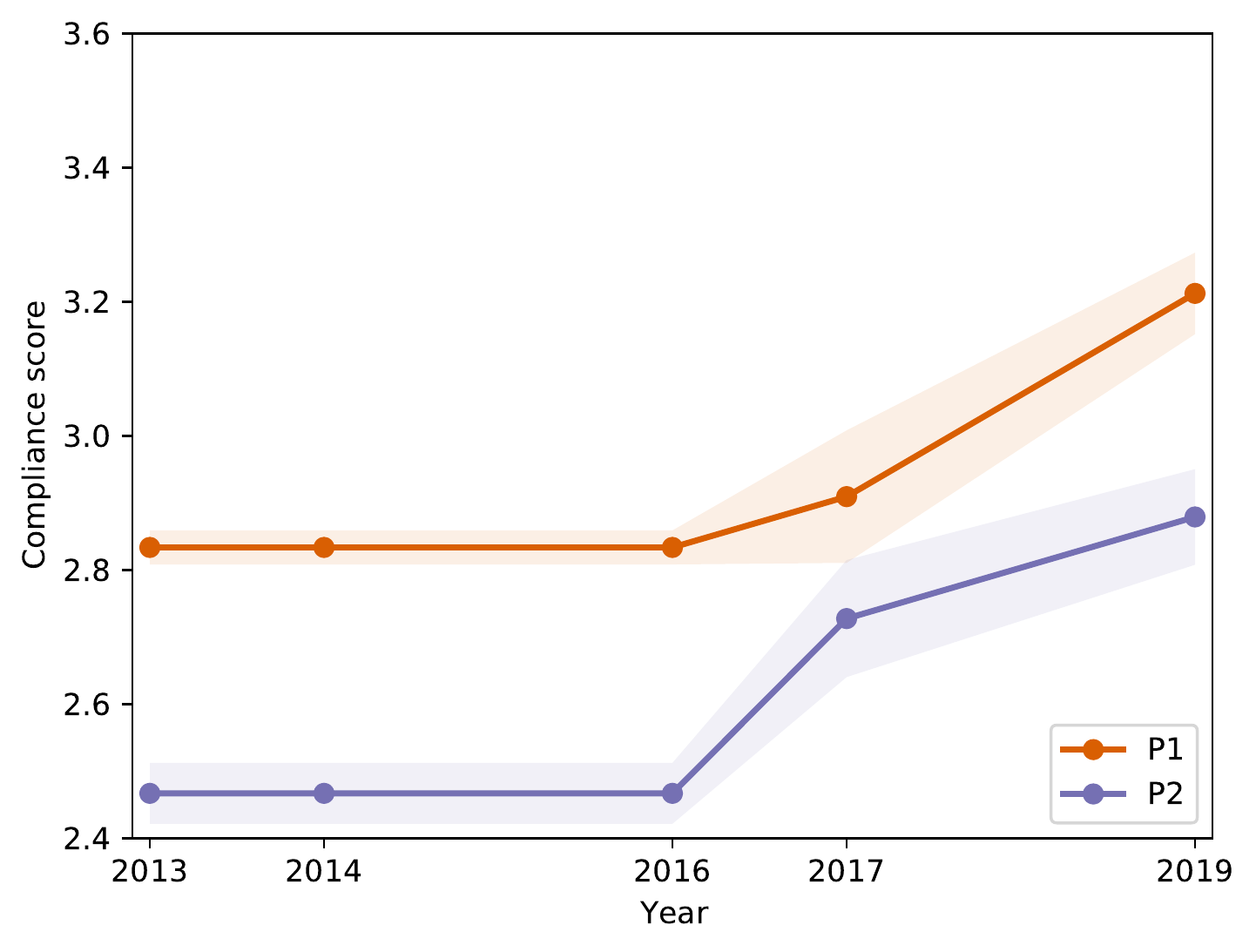}
\includegraphics[width=0.33\textwidth,keepaspectratio]{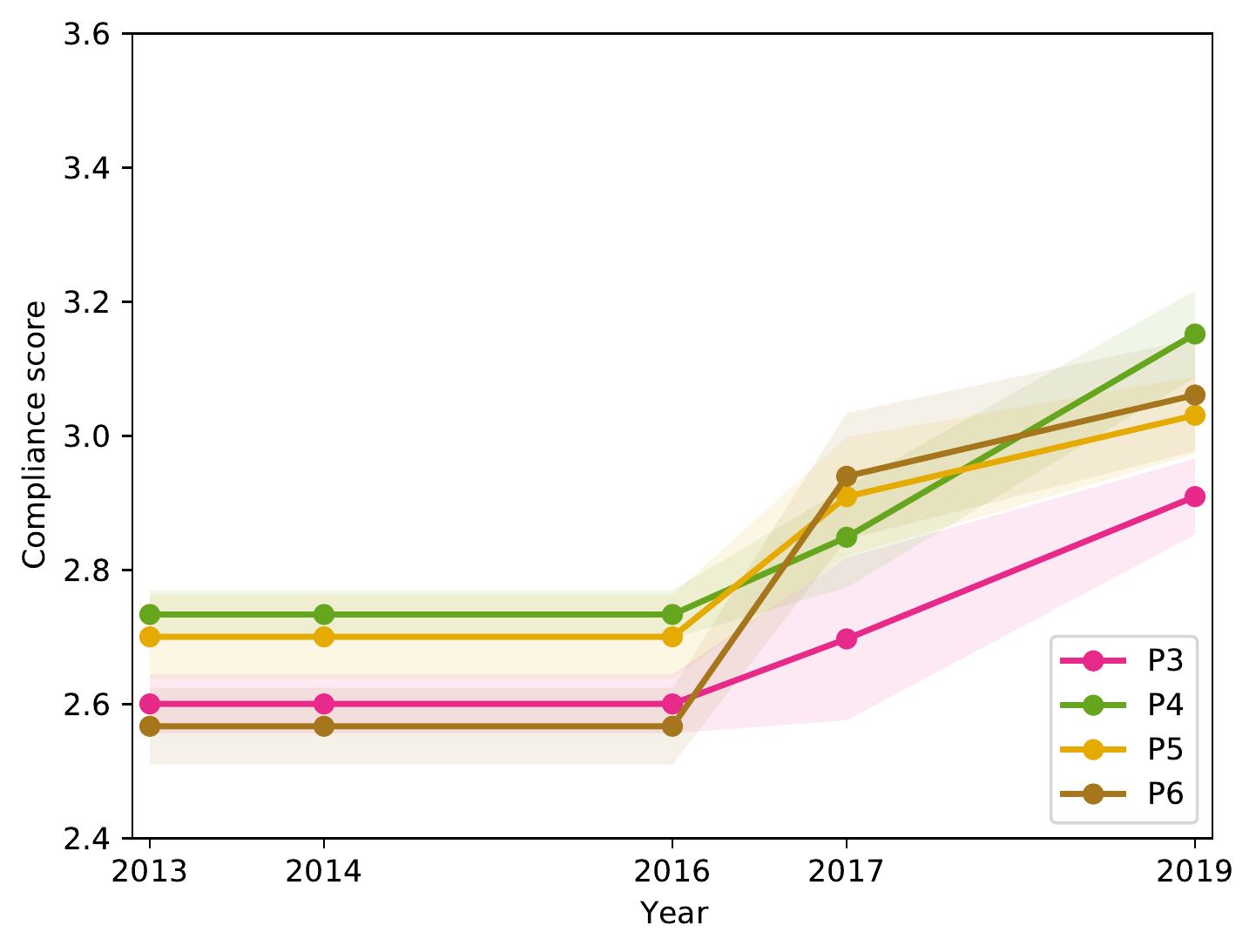}
\includegraphics[width=0.33\textwidth,keepaspectratio]{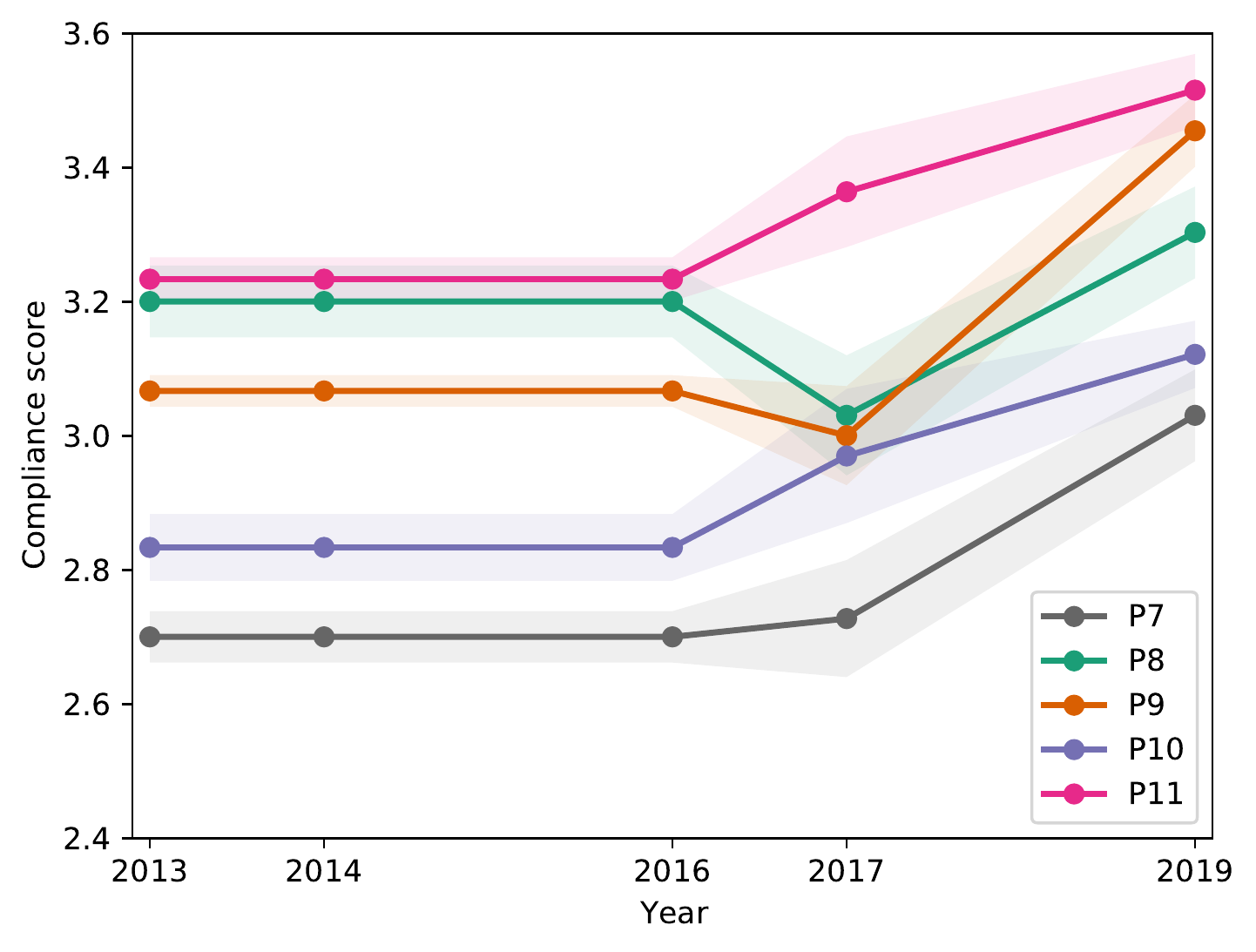}
\caption{
Worldwide compliance with BCBS 239 has missed the original 2016 deadline.
Shown are the mean compliance scores (1 = not compliant, 4 = fully compliant) across the evaluated banks, together with the standard error in the light shaded envelopes.
While compliance with Topics I and II (left and center respectively) have in general improved,
compliance with the principles of Topic III (right) have for the most part stayed stagnant.
}
\label{fig:compliance-scores}
\end{figure*}

These 11 principles form the core of periodic compliance assessments carried out by BCBS, which were conducted on G-SIBs in the years 2013, 2014, 2016, and 2017 \cite{bcbs239pr}.
\Cref{fig:compliance-scores} shows the mean compliance scores from the progress reports, on a four-point Likert scale (1 = not compliant, 4 = fully compliant) across the evaluated banks, together with the standard error in the light shaded envelopes.
Overall, worldwide compliance with BCBS 239 among G-SIBs has missed the original 2016 deadline.
While compliance with Topics I and II (left and center respectively) have in general improved,
compliance with the principles of Topic III (right) have for the most part stayed stagnant, and even shown some regression (P8, P9).
Widening standard errors from 2016 to 2017 also indicate a widening gaps in compliance levels between individual banks.

The progress reports note that one of the key difficulties in compliance lies in automation.
However, we now offer some statistical evidence in favor of a claim that explainable AI is
needed to make progress in BCBS compliance.
\Cref{tab:compliance-change} 
reports mean compliance scores in the years 2016 and 2017 (with standard error in parentheses),
the improvement from 2016 to 2017,
and the $p$ value for the one-sided $t$-test for the hypothesis test that the mean compliance score improved in 2017 relative to 2016.
Data is from the annual BCBS progress reports.
We use the pooled sample (independent) $t$-test because scores for each bank are not publicly available.
Values that do not meet the usual $p<0.05$ criterion are \textit{italicized}.
The largest three values are in \textbf{\textit{bold italics}}.
The last column shows our ratings of the principle's relevance to explainable AI (* = minor, ** = major).
The top three principles that show the least evidence for improvement (by $p$-value)
are exactly those most strongly related to explainable AI needs.

\begin{table}
\centering
{\small
\begin{tabular}{|c|c|c|c|c|c|}
\hline 
    & 2017          & 2019          &$\Delta$       & $\nu$& $p$   \tabularnewline
\hline
\input{bcbs_table_rows}
\hline
\end{tabular}
}

\caption{Lack of compliance progress is related to explainable AI needs.
Reported are mean compliance scores in the years 2016 and 2017 (with standard error in parentheses),
the improvement from 2016 to 2017,
and the $p$ value for the one-sided $t$-test for the hypothesis test that the mean compliance score improved in 2017 relative to 2016.
Values that do not meet the usual $p<0.05$ criterion are \textit{italicized}.
The largest three values are also in \textbf{\textit{bold}}.
The last column shows our ratings of the principle's relevance to explainable AI (* = minor, ** = major).
The top three principles that show the least evidence for improvement (by $p$-value)
are exactly those most strongly related to explainable AI needs.
}
\label{tab:compliance-change}
\end{table}

From our qualitative analysis in the previous section, 
note that Principles P6, P7 and P9 are the principles most closely aligned with explainability needs, and these are precisely the principles that show the largest decline in compliance.

In this discussion, we look at data quality as a necessary component of explainable AI.
In particular, we look the challenge of building a coherent data taxonomy.

\section{Metadata debt in legacy enterprises}

Enterprises that collect data eventually have to face the challenges of data governance and data management \cite{Khatri2010,Redman2013}.
In some heavily regulated industries, such as financial services, having good data management is even a matter of regulatory compliance \cite{bcbs239}.
In this section, we consider the problem of \textit{metadata debt},
a kind of technical debt incurred by legacy enterprises when they lack a unified ontology for describing the various kinds of data that they have \cite{Chen2020}.
A key aspect of good data governance is to develop have metadata that describe the semantic content of data in a way that is interpretable to end users \cite{Khatri2010,robcasper}.
However, retrofitting good data governance onto legacy enterprises is a major challenge,
especially when the infrastructure necessary to track the creation and consumption of data does not exist.

Legacy enterprises may attempt to pay down metadata debt with a generalized form ontology learning, deriving a common vocabulary of data concepts not just from natural text, but also other sequence structures present in relational databases, log files, and other forms of metadata.
When doing so, they will face several practical challenges:

\paragraph{Indirect and polymorphic representation}
The same kind of data can have many physical representations, be they structured or unstructured,
in databases, flat files, text, images, or other binary blobs.
Even when structured representations are used,
there may be many such representations for data of the same semantics.
Such polymorphism can result from the merger of multiple legacy systems that use incompatible representations,
which nevertheless have to be harmonized to provide a complete representation of the available data.
Having an accurate and precise relationship between semantic and physical data is crucial for other aspects of data governance, such as data access, data quality, and lifecycle management.

\paragraph{High cost, high accuracy requirements despite label noise}
The creation of semantic labels is very expensive, yet high accuracy is required.
The high cost comes from the need to rely on subject matter experts to provide the human labor to label data semantically,
as well as to verify the correctness of existing labels.
The need for high quality distinguishes this problem to many deep learning applications such as image classification, where labels are crowdsourced, easily verifiable, and essentially free to acquire.
High accuracy requires us to detect and mitigate label noise, but conventional methods such as explicit testing for inter-rater reliability are usually too expensive to run.
There is therefore a need for a method that automatically quantifies label noise.

\paragraph{Large controlled vocabulary}
The size of the controlled vocabulary for the semantic metadata can be itself large for a large organization with multiple kinds of data.
Terms in the controlled vocabulary are rarely, if ever, used with equal frequency---there will be some frequently used terms,
accompanied by a tail of many, successively infrequent, terms.
The existence of many rare terms makes it difficult for the na\"ive approach of training many independent classifiers,
simply because there are insufficient positive examples of usage for the rare terms: the low signal-to-noise ratio makes it difficult to train a classifier with performance better than random.

\paragraph{Organizational changes}
The high cost of changing systems in production create an incentive to reappropriate existing data systems for new tasks and new representations of data, but the same attitude of cost avoidance also means that the costly task of updating semantic metadata is often skipped over.
In other situations, the organization changes such as restructuring may result in a loss of subject matter experts which are conversant in the informal folklore,
which further raises the difficulty of acquiring labels.

\paragraph{Ontology drift and concept drift}
The physical representations of a data concept can change over time, as APIs and data formats change \cite{Gama2014}.
Furthermore, the meaning of the concept itself can change over time due to changes in usage \cite{Wang2011,Kenter2015}.
Even seemly unchanging ontologies like the Dewey Decimal Classification for library books have undergone 23 major print editions in the years 1876---2011, and is now continuously updated \cite{ddc}.
There is therefore a need for an ontology learning method can can be rerun periodically to detect such concept drifts and ontology drifts.

\subsection{There are multiple relevant industry taxonomies and ontologies}

We now examine one specific requirement in greater detail:

\begin{displayquote}
33. A bank should establish integrated data taxonomies and architecture across the
banking group, which includes information on the characteristics of the data (metadata), as
well as use of single identifiers and/or unified naming conventions for data including legal
entities, counterparties, customers and accounts. [...] Banks do not necessarily need to have one data model; rather, there should be robust automated reconciliation procedures where multiple models are in use.
\end{displayquote}

In practice, there are multiple business-relevant taxonomies and ontologies for the financial services industry.
An incomplete list of these are:

\begin{itemize}
\item
GCIS \cite{gics},
to classify companies by industrial sectors.

\item
Solvency II DPM \cite{eiopa-dpm},
describing business concepts relevant to solvency testing.

\item
IFRS Taxonomy \cite{ifrs},
which describes financial statements.

\item
ESEF Taxonomy \cite{esef},
used for electronic reporting within the European Union.

\item
US GAAP Taxonomy \cite{us-gaap},
which is used for accounting within the US.

\item
LEI Taxonomy \cite{lei},
which describes unique global identifiers for legal entities participating in financial transactions.

\item
FRC Taxonomy \cite{frc},
which is used for accounting within the UK.

\item
BIRD \cite{ecb-bird},
which describes information in bank internal systems
for fulfilling their reporting requirements within the European System of Central Banks.


\item
FIBO \cite{Bennett2013,fibo},
a general purpose ontology for all business concepts in the financial industry.

\end{itemize}

Each taxonomy has its own roadmap and cadence of updates. Some, like FIBO, are updated quarterly. Others, like IFRS, are updated annually. Yet others have more infrequent or irregular schedules of updates.

Furthermore, the taxonomies come in different formats which have to be integrated. While many of the taxonomies above are in the industry-standard XML-based XBRL \cite{xbrl}, some are still published as Microsoft Excel spreadsheets (like NAICS \cite{naics}), and others are published as Microsoft Access databases (like BIRD). Therefore there is also the technical challenge of interoperability between different formats to consider.

\subsection{Changes in an industry standard ontology}

\begin{figure}
\centering
\includegraphics[width=\columnwidth]{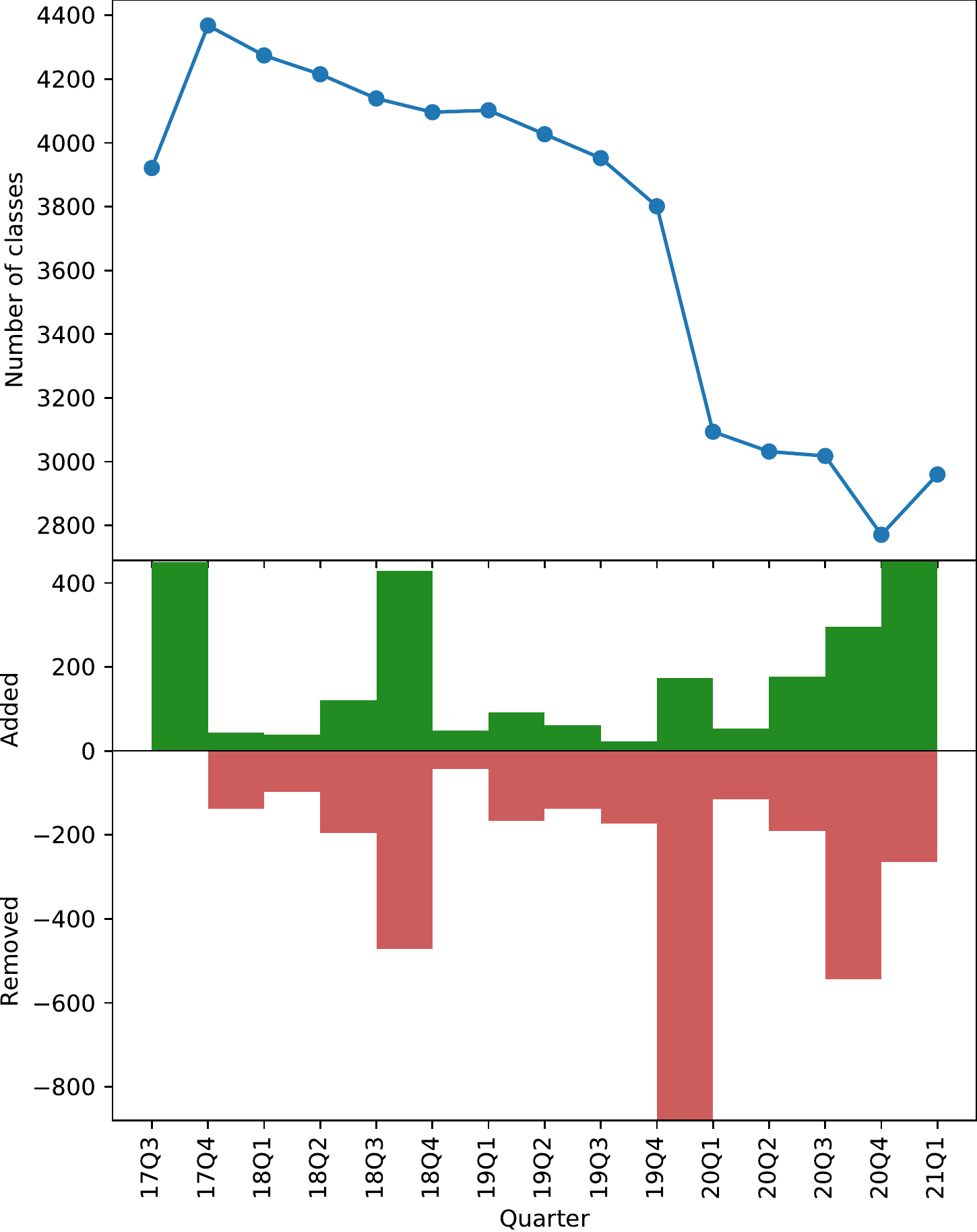}
\caption{Changes in FIBO classes}
\label{fig:fibo}
\end{figure}

The Financial Industry Business Ontology (FIBO) project \cite{Bennett2013,fibo}  was launched in the same year that BCBS 239 was published.
The first production-ready released was in 2017Q3, with a quarterly update cadence.
Shown in \Cref{tab:fibo,fig:fibo} are the changes in the list of classes (data concepts) up to the most recently available version.
The results presented show that the business ontology is in constant flux, with as much as one third of the entire ontology changing quarterly.
Rather than converging toward a steady state, the pace of change seems erratic.
These simple statistics show that mapping onto an industry standard ontology is far from a straightforward, one-time exercise;
rather, changes in the ontology have to be versioned and managed as concepts are added, redefined, or removed.
Explainable AI work that aims to map onto FIBO or other standardized ontologies must therefore heed the ensuing challenges of ontology drift
and other semantic changes.

\begin{table}
\centering
\begin{filecontents*}{data/fibo-changes.tsv}
Quarter,No. concepts,Added,Removed,Change
2017Q3,3921,,,
2017Q4,4368,+449,-2,10.3\%
2018Q1,4274,+44,-138,4.3\%
2018Q2,4215,+38,-97,3.2\%
2018Q3,4139,+120,-196,7.6\%
2018Q4,4096,+429,-472,22.0\%
2019Q1,4102,+49,-43,2.2\%
2019Q2,4027,+92,-167,6.4\%
2019Q3,3952,+62,-137,5.0\%
2019Q4,3801,+22,-173,5.1\%
2020Q1,3094,+173,-880,34.0\%
2020Q2,3032,+54,-116,5.6\%
2020Q3,3018,+176,-190,12.1\%
2020Q4,2771,+296,-543,30.3\%
2021Q1,2960,+454,-265,24.3\%
\end{filecontents*}
\csvautotabular{data/fibo-changes.tsv}
\caption{Changes in the FIBO ontology}
\label{tab:fibo}
\end{table}

\section{Summary and outlook}

We have offered our interpretation of BCBS 239 as encoding 8 distinct needs for explainable AI,
which we grouped into data quality, and appropriate reporting for multiple stakeholders.
Evidence from compliance progress reports demonstrates that the areas with the slowest progress
toward compliance is also the same areas that are most strongly related to explainable AI.

We also took a closer look at the construction and maintenance of a firmwide data taxonomy,
being one of the needs for explainable AI that I have identified.
We described the implementation challenges for a specific requirement that occur in legacy
enterprises that have to retrofit their business operations and infrastructure to support
a firmwide data taxonomy.
Finally, we reviewed how a candidate standardized solution to this problem,
namely the Financial Industry Business Ontology (FIBO),
highlights the need for versioning and updating to prevent semantic drift.
We therefore expect explainable AI to be a constant need for the foreseeable future
when it comes to standards compliance,
and that systems for representing explanations must handle semantic changes
to avoid obsolescence in the concepts themselves.

\paragraph{Disclaimer}
This paper was prepared for informational purposes by the Artificial Intelligence Research  group of JPMorgan Chase \& Co and its affiliates (``JP Morgan''), and is not a product of the Research Department of JP Morgan. JP Morgan makes no representation and warranty whatsoever and disclaims all liability, for the completeness, accuracy or reliability of the information contained herein.  This document is not intended as investment research or investment advice, or a recommendation, offer or solicitation for the purchase or sale of any security, financial instrument, financial product or service, or to be used in any way for evaluating the merits of participating in any transaction, and shall not constitute a solicitation under any jurisdiction or to any person, if such solicitation under such jurisdiction or to such person would be unlawful. © 2021 JPMorgan Chase \& Co. All rights reserved.

\bibliographystyle{ACM-Reference-Format}
\bibliography{bib}

\end{document}

%% file: bcbs_table_rows.tex
 P1 & 2.909 (0.099) & 3.212 (0.061) & 0.303 (0.116) & 53.2 & 0.382 \tabularnewline 
 P2 & 2.727 (0.087) & 2.879 (0.071) & 0.152 (0.113) & 61.5 & 0.440 \tabularnewline 
\hline
 P3 & 2.697 (0.121) & 2.909 (0.057) & 0.212 (0.134) & 45.3 & 0.416 \tabularnewline 
 P4 & 2.848 (0.075) & 3.152 (0.065) & 0.303 (0.099) & 62.6 & 0.381 \tabularnewline 
 P5 & 2.909 (0.088) & 3.030 (0.058) & 0.121 (0.106) & 55.2 & 0.452 \tabularnewline 
 P6 & 2.939 (0.094) & 3.061 (0.084) & 0.121 (0.126) & 63.1 & 0.452 \tabularnewline 
\hline
 P7 & 2.727 (0.087) & 3.030 (0.068) & 0.303 (0.111) & 60.5 & 0.381 \tabularnewline 
 P8 & 3.030 (0.090) & 3.303 (0.068) & 0.273 (0.113) & 59.9 & 0.393 \tabularnewline 
 P9 & 3.000 (0.074) & 3.455 (0.054) & 0.455 (0.091) & 58.5 & 0.326 \tabularnewline 
P10 & 2.970 (0.100) & 3.121 (0.050) & 0.152 (0.112) & 47.1 & 0.440 \tabularnewline 
P11 & 3.364 (0.082) & 3.515 (0.054) & 0.152 (0.099) & 55.2 & 0.440 \tabularnewline 